V.I. Abrosimov[*], A.I. Levon

*Institute for Nuclear Research, National Academy of Sciences of Ukraine*

[*]Corresponding author: abrosim@kinr.kiev.ua

# EXCITATION OF MONOPOLE PAIRING VIBRATIONS IN TWO-NEUTRON TRANSFER REACTION: A SEMICLASSICAL APPROACH

For studying the collective pairing excitations of nuclei, the two-nucleon transfer reactions in superfluid nuclei (the pairing gap of the ground state is not zero), in particular, the (p,t) reaction, are of the greatest interest. A simple model of monopole pairing excitations in superfluid nuclei on the basis of the semiclassical time-dependent Hartree-Fock-Bogolyubov theory in the limit of small amplitudes is considered. Using the anomalous density response function, the monopole pairing mode in the energy region of double pairing gap and the variation of the pairing gap associated with this mode are found. The ratio of the spectroscopic factor for the excitation of monopole pairing vibrations in the (p,t) reaction in even superfluid nuclei to the spectroscopic factor for the transfer of two neutrons to the ground state of the daughter nucleus (the relative spectroscopic factor) is estimated. For this, it is assumed that the relative spectroscopic factor is proportional to the pairing gap variation associated with the monopole pairing vibrations in agreement with the corresponding quantum expression. Numerical estimate of the relative spectroscopic factors for superfluid nuclei of the rare-earth and actinide regions shows that the spectroscopic factor for the two-neutron transfer, leading to the excitation of the monopole pairing vibrations, does not exceed several percent of the spectroscopic factor for the transfer of two neutrons to the ground state. This semiclassical estimate is in agreement with the experimental rations of the cross sections for the excitation of $0^+$ states in the energy region of double pairing gap to the cross sections for the excitation of the ground states for superfluid nuclei of the rare-earth and actinide regions.

## 1. Introduction

Recent experimental studies of the two-neutron transfer reaction on superfluid nuclei (deformed nuclei of rare-earth and actinide regions) have shown many new excited states in the low-energy region (up to 4.3 MeV), especially a lot of new $0^+$ states [1-3]. It is known that the two-neutron transfer reaction is an effective tool for searching of collective effects associated with pairing correlations in nuclei [4,5]. It is of interest to look for pairing effects in this new experimental information, in particular, purely pairing vibrations associated with dynamic fluctuations of pairing gap. Theoretical studies of pairing vibrations in nuclei has been one of important topic for many years [6-11] and different theoretical approaches have been developed (see [12,13] and references therein). It is reasonable to study purely pairing vibrations related to dynamic variations of pairing gap using a semiclassical approach that allows us to describe the average properties of the nuclear pairing vibrations in a physically transparent way.

In present paper, we consider excitation of monopole pairing vibrations in superfluid nuclei in the two-neutron transfer reaction within an approach based on the semiclassical time-dependent Hartree-Fock-Bogolyubov theory (extended Vlasov kinetic equation with pairing) [14,15]. In Section 2, we briefly recall the self-consistent model of pairing vibrations, which uses the semiclassical time-dependent Hartree-Fock-Bogolyubov equations of motion for small amplitudes. To study the monopole pairing excitations in superfluid nuclei the anomalous (correlated) density response function is considered. In Section 3, the relative spectroscopic factor for the excitation of monopole pairing vibrations in the two-neutron transfer reaction is estimated. Numerical evaluations of the relative spectroscopic factor are compared with the corresponding experimental relative cross sections for the excitation of $0^+$ states in the (p,t) reaction in superfluid nuclei of the rare-earth and actinide regions [1-3].

## 2. Semiclassical model of pairing vibrations

We consider a simple model of the collective pairing excitations in superfluid finite Fermi systems, which is based on the semiclassical equations of motion of the time-dependent Hartree-Fock-Bogoliubov theory for small amplitudes [15]. In semiclassical pairing theory, the spectrum of eigenfrequencies has a pairing gap in both deformed and spherical finite Fermi systems, since quantum effects (shell effects) are not included in this theory. We will use this property to simplify the model.

To study the collective pairing excitations, we consider the dynamical equation for the anomalous phase-space distribution function

$$i\hbar \partial_t \delta\kappa = 2(h_0 - \mu)\delta\kappa - (2\rho_0 - 1)\delta\Delta + 2\kappa_0 \delta h - 2\Delta_0 \delta\rho_{ev} \tag{1}$$

with

$$\delta\kappa(\mathbf{r},\mathbf{p},t) = \delta\kappa_r(\mathbf{r},\mathbf{p},t) + i\delta\kappa_i(\mathbf{r},\mathbf{p},t). \tag{2}$$

We have the equation of motion for the variation of the anomalous phase-space distribution function $\delta\kappa(\mathbf{r},\mathbf{p},t)$ from equilibrium distribution $\kappa_0(\mathbf{r},\mathbf{p})$. The variation $\delta\kappa(\mathbf{r},\mathbf{p},t)$ is a complex function. To first order, $\delta\kappa_r(\mathbf{r},\mathbf{p},t)$ gives the change in magnitude of $\kappa(\mathbf{r},\mathbf{p},t)$ and $\delta\kappa_i(\mathbf{r},\mathbf{p},t)$ is proportional to the change in the phase of $\kappa(\mathbf{r},\mathbf{p},t)$. The variation of the even normal phase-space distribution function $\delta\rho_{ev}(\mathbf{r},\mathbf{p},t)$ from equilibrium distribution $\rho_0(\mathbf{r},\mathbf{p})$ is a real function. In order to determine this quantity, we need an extra equation for $\delta\rho_{ev}(\mathbf{r},\mathbf{p},t)$. We use a supplementary condition enforced by the Pauli principle [16]. It reads

$$(2\rho_0 - 1)\delta\rho_{ev}(\mathbf{r},\mathbf{p},t) + 2\kappa_0 \delta\kappa_r(\mathbf{r},\mathbf{p},t) = 0. \tag{3}$$

The variation of the pairing field $\delta\Delta(\mathbf{r},\mathbf{p},t)$ from equilibrium value $\Delta_0(\mathbf{r},\mathbf{p})$ is a complex function. We want to study the solution of dynamical equation (1) for finite Fermi systems taking into account the self-consistent pairing-field fluctuations $\delta\Delta(\mathbf{r},\mathbf{p},t)$ related to the residual pairing interaction. To achieve this, we assume that the variation of the pairing field $\delta\Delta(\mathbf{r},\mathbf{p},t)$ is associated with the variation of the anomalous phase-space distribution $\delta\kappa(\mathbf{r},\mathbf{p},t)$ by the gap equation of the BCS type [14]. Then we get the self-consistency condition:

$$\int \frac{d\mathbf{p}}{(2\pi\hbar)^3}\left(\delta\kappa(\mathbf{r},\mathbf{p},t) - \kappa_0(\mathbf{r},\mathbf{p})\frac{\delta\Delta(\mathbf{r},t)}{\Delta_0(\mathbf{r},\mathbf{p})}\right) = 0. \tag{4}$$

It can be seen that the real part of pairing field variation $\delta\Delta_r(\mathbf{r},\mathbf{p},t)$ gives the change in magnitude of the equilibrium pairing gap. We get two additional equations for $\delta\Delta_{r,i}(\mathbf{r},\mathbf{p},t)$ and a closed system of the dynamical equations. In the following, we approximate the equilibrium pairing field $\Delta_0(\mathbf{r},\mathbf{p})$ with the phenomenological parameter $\Delta$.

The two equilibrium phase-space distributions $\rho_0(\mathbf{r},\mathbf{p})$ and $\kappa_0(\mathbf{r},\mathbf{p})$ in Eqs. (1), (3) and (4) are given by [17]

$$\rho_0(\mathbf{r},\mathbf{p}) = \frac{1}{2}\left(1 - \frac{h_0(\mathbf{r},\mathbf{p}) - \mu}{E(\mathbf{r},\mathbf{p})}\right), \tag{5}$$

$$\kappa_0(\mathbf{r},\mathbf{p}) = -\frac{\Delta_0(\mathbf{r},\mathbf{p})}{2E(\mathbf{r},\mathbf{p})} \tag{6}$$

with the quasiparticle energy

$$E(\mathbf{r},\mathbf{p}) = \sqrt{(h_0(\mathbf{r},\mathbf{p}) - \mu)^2 + \Delta_0^2(\mathbf{r},\mathbf{p})}. \tag{7}$$

The chemical potential $\mu$ is determined by the condition

$$A = \frac{4}{(2\pi\hbar)^3}\int d\mathbf{r}\, d\mathbf{p}\, \rho_0(\mathbf{r},\mathbf{p}), \tag{8}$$

where $A$ is the number of nucleons in the system. The equilibrium single-particle Hamiltonian

$$h_0(\mathbf{r},\mathbf{p}) = \frac{p^2}{2m} + V_0(\mathbf{r}) \tag{9}$$

contains the self-consistent mean field $V_0(\mathbf{r})$ however in the following we approximate the static nuclear mean field with a spherical square-well potential of radius $R$. This choice allows us to take into account finite-size effects and, at the same time, to recover the simplicity of homogeneous systems: the static and dynamic equations become functions of the particle energy $\epsilon = h_0(\mathbf{r},\mathbf{p})$ alone.

We shall consider the zero-order approximation for the normal mean field, neglecting self-consistent variations of the normal mean field ($\delta h(\mathbf{r},t) \approx 0$ in Eq. (1)).

Eqs. (1), (3) and (4) are a closed system of the dynamical equations for the anomalous phase-space distributions $\delta\kappa_{r,i}(\mathbf{r},\mathbf{p},t)$ that we have to solve. We will consider the monopole vibrations related to the self-consistent pairing-field fluctuations, thus we assume that the anomalous density fluctuations are induced by the monopole external field of the kind

$$U^{ext}(r,t) = \beta\, \delta(t)\, f(r), \tag{10}$$

where $f(r) = \Theta(r-R)$ and $\beta$ is a small parameter specifying the strength of the external field. We assume that the external field causes the extra fluctuations of the real part of the pairing field $\delta\Delta_r(\mathbf{r},t)$. Thus the fluctuations $\delta\Delta_r(\mathbf{r},t)$ in Eq. (1) are treated as

$$\delta\Delta_r(r,t) = \delta\Delta_r^{int}(r,t) + U^{ext}(r,t), \tag{11}$$

where $\delta\Delta_r^{int}(r,t)$ are the self-consistent fluctuations due to the residual pairing interaction.

Taking into account our approximations and using Eqs. (5), (6) gives the considerably simplified system of coupled equations. Their time Fourier transform is given by

$$-i\hbar\omega\, \delta\kappa_i(r,\epsilon,\omega) = 2(\epsilon-\mu)\left[\delta\kappa_r(r,\epsilon,\omega) + \frac{\delta\Delta_r(r,\omega)}{2E(\epsilon)}\right] - 2\Delta\delta\rho_{ev}(r,\epsilon,\omega), \tag{12}$$

$$i\hbar\omega\, \delta\kappa_r(r,\epsilon,\omega) = 2(\epsilon-\mu)\left[\delta\kappa_i(r,\epsilon,\omega) + \frac{\delta\Delta_i(r,\omega)}{2E(\epsilon)}\right], \tag{13}$$

$$(\epsilon-\mu)\delta\rho_{ev}(r,\epsilon,\omega) + \Delta\delta\kappa_r(r,\epsilon,\omega) = 0, \tag{14}$$

$$\int d\epsilon\, g(\epsilon)\left(\delta\kappa_{r,i}(r,\epsilon,\omega) + \frac{\delta\Delta_{r,i}^{int}(r,\omega)}{2E(\epsilon)}\right) = 0. \tag{15}$$

Here $g(\epsilon)$ is the pair-neutron level density per unit volume in the equilibrium mean field, which is approximated in our model by a spherical square-well potential, so we can get

$$g(\epsilon) = \frac{1}{4\pi^2}\left(\frac{2m}{\hbar^2}\right)^{3/2}\sqrt{\epsilon}. \tag{16}$$

By using the system of coupled equations (12)-(15), we can find the expression for the anomalous density variation $\delta\kappa_r(r,\omega)$ induced by the monopole external field (10) and define the monopole anomalous density response function as

$$R_{PV}(r,\omega) = \frac{1}{\beta}\frac{2}{(2\pi\hbar)^3}\int d\mathbf{p}\, \delta\kappa_r(r,\mathbf{p},\omega) = \frac{1}{\beta}\, \delta\kappa_r(r,\omega). \tag{17}$$

The anomalous density variation $\delta\kappa_r(r,\omega)$ is given by [15]

$$\delta\kappa_r(r,\omega) = \frac{\alpha R_r^0(\omega)}{\alpha + R_r^0(\omega)} U^{ext}(r,\omega) \qquad (18)$$

with

$$R_r^0(\omega) = I_3(\omega) - \frac{[I_2(\omega)]^2}{I_1(\omega)}, \quad \alpha = \int_0^{\epsilon_c} d\epsilon\, g(\epsilon) \frac{1}{E(\epsilon)}, \qquad (19)$$

where

$$I_i(\omega) = \int_0^{\epsilon_c} d\epsilon\, g(\epsilon) \frac{f_i(\epsilon)}{E^2(\epsilon)} \left[ \frac{1}{\hbar\omega - 2E(\epsilon) + \iota\eta} - \frac{1}{\hbar\omega + 2E(\epsilon) + \iota\eta} \right] \qquad (20)$$

with $f_1(\epsilon)=1$, $f_2(\epsilon)=\epsilon-\mu$ and $f_3(\epsilon)=(\epsilon-\mu)^2$.

The results of numerical calculations of the strength function associated with the monopole anomalous density response function (17) as

$$S_{PV}(\hbar\omega)/\alpha = -\frac{1}{\pi} \mathrm{Im}\, R_{PV}(\hbar\omega)/\alpha \qquad (21)$$

are shown in Fig.1 ($E = \hbar\omega$). In our calculations we used the standard values of nuclear parameters:

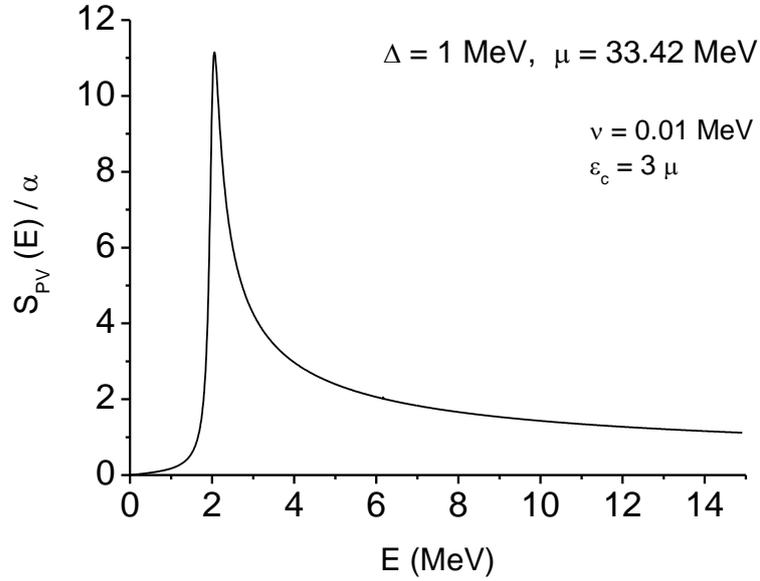

*Fig.1. The strength function of the monopole anomalous density for finite system of correlated nucleons taking into account the residual pairing interaction.*

$r_0 = 1.2$ fm, $\mu \approx \epsilon_F = 33.42$ MeV, $m = 1.04$ MeV$(10^{-22}$s$)^2$/fm$^2$, $\Delta = 1$ MeV. The strength function has a resonance structure with a sharp peak around $2\Delta$ that display the monopole collective pairing mode. The width of this mode is due to the Landau damping. In Fig. 2 the dependence of the strength function on the parameter of particle energy cut off $\epsilon_c$ is shown. It is seen that the position and width of the monopole pairing resonance change slightly depending on the value of the particle energy cut off parameter.

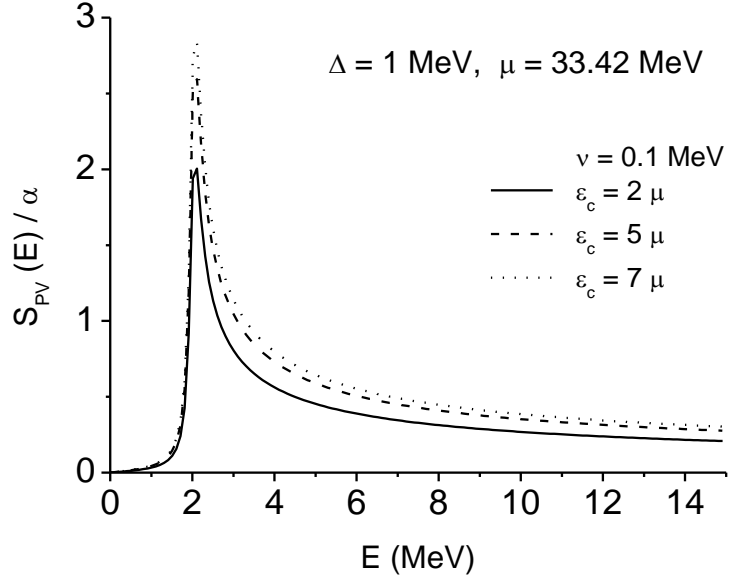

*Fig.2. The dependence of the monopole anomalous strength function on the parameter of particle energy cut off $\epsilon_c$.*

In the next section, we use the monopole anomalous density response function (16) to estimate the spectroscopic factor for the two-neutron transfer, leading to the excitation of the monopole pairing vibrations in superfluid nuclei.

### 3. Spectroscopic factor for two-neutron transfer reaction

Pure pairing collective excitations associated with dynamic fluctuations of the pairing gap can reveal itself in the reactions of two neutron transfer in superfluid nuclei. In order to evaluate the intensity of the excitation of pairing vibrations in these reactions, it is convenient to consider the ratio of the spectroscopic factor for the two-neutron transfer, which leads to the excitation of pairing vibrations, to the spectroscopic factor for the transfer of two neutrons to the ground state (the relative spectroscopic factor). This quantity is usually used in analyses of the experimental data.

It was found in Ref. [18] (see Eq. (49)) that the relative spectroscopic factor for the two-neutron transfer is proportional to the variation of the pairing gap in the dynamical system. In analogy with the quantum estimate, we will assume that the semiclassical relative spectroscopic factor for the two-neutron transfer $S_{PV}(p,t)/S_0(p,t)$ is determined by the squared amplitude of the pairing gap variation $\delta\Delta_r(\omega_{PV})$ associated with the monopole pairing vibrations as

$$\frac{S_{PV}(p,t)}{S_0(p,t)} \approx \frac{|\delta\Delta(\omega_{PV})|^2}{\Delta^2}. \tag{22}$$

We use the monopole anomalous density response function (16) that is related to the dynamical fluctuations of the pairing field to estimate the squared amplitude of the pairing gap vibrations. The poles of the response function (16) determine the pairing collective modes (pairing vibrations) that are given by the roots of vanishing denominator of the anomalous density variation $\delta\kappa_r(r,\omega)$:

$$\alpha + R_r^0(\omega) = 0. \tag{23}$$

Using Eqs. (19), (20), this equation can be rewritten in the form

$$[(\hbar\omega)^2 - 4\Delta^2]\frac{I_1(\omega)}{4} - \frac{[I_2(\omega)]^2}{I_1(\omega)} = 0. \tag{24}$$

Taking into account that $|I_2(\omega)/(\Delta\, I_1(\omega))| \ll 1$, one can get the energy of the monopole pairing mode at $\hbar\omega_{pv} \approx 2\Delta$. The equation for the eigenfrequencies (24) is a semiclassical analogue of the corresponding quantum equation [12]. In the quantum theory, integrals in Eq. (24) are replaced by sums over discrete quantum levels.

The eigenfrequency equation (24) can be interpreted as related to the harmonic oscillator describing the vibrations of the pairing gap near the equilibrium value $\Delta$ in finite systems. It can be written as

$$\omega^2 B(\omega) - C(\omega) = 0. \tag{25}$$

Here the kinetic parameters $B(\omega)$ and $C(\omega)$ are defined for finite spherical systems as

$$B(\omega) = \frac{\hbar^2}{4}\int d\vec{r}\, |I_1(\omega)| = \frac{\hbar^2}{4}|\tilde{I}_1(\omega)|, \tag{26}$$

$$C(\omega) = \Delta^2 |\tilde{I}_1(\omega)| + \frac{[\tilde{I}_2(\omega)]^2}{|\tilde{I}_1(\omega)|} \tag{27}$$

with

$$\tilde{I}_i(\omega) = \int_0^{\epsilon_c} d\epsilon\, \tilde{g}(\epsilon)\frac{f_i(\epsilon)}{E^2(\epsilon)}\left[\frac{1}{\hbar\omega - 2E(\epsilon) + \iota\eta} - \frac{1}{\hbar\omega + 2E(\epsilon) + \iota\eta}\right] \tag{28}$$

where $\tilde{g}(\epsilon) = (4\pi/3)R^3 g(\epsilon)$ is the energy density for a spherical square-well potential of radius $R$.

To evaluate the amplitude of the pairing gap vibrations, we use the expression for the energy of harmonic oscillator describing the pairing gap vibrations and "the quantum relation" between this semiclassical energy and eigenfrequency of the pairing vibrations. Then we get

$$\omega_{PV}^2 \, B(\omega_{PV}) \, |\delta\Delta(\omega_{PV})|^2 = \hbar\omega_{PV} \tag{29}$$

and

$$|\delta\Delta(\omega_{PV})|^2 = \frac{\hbar\omega_{PV}}{\omega_{PV}^2 \, B(\omega_{PV})}. \tag{30}$$

Here the parameter $B(\omega_{PV})$ at $\omega_{PV} \approx 2\Delta/\hbar$ is given by

$$B(\omega_{PV}) = \frac{\hbar^2}{4} \int_0^\infty d\epsilon \, \tilde{g}(\epsilon) \frac{1}{E(\epsilon)(\epsilon-\mu)^2}. \tag{31}$$

It is seen that the parameter $B(\omega_{PV})$ contains the coherence, which is essentially determined by the distribution of the neutron levels near the Fermi energy. The similar coherence related to the pairing excitations in two-neutron reaction was found in quantum approaches [6,19]. The semiclassical expression (31) has the second order pole at $\epsilon=\mu$, so in order to estimate this quantity a more accurate approximation is needed for the distribution of the single-particle energy near the Fermi energy. Of course, the details of the level density near the Fermi energy in finite Fermi systems are determined by quantum effects. However, since the present semiclassical approach leads to rather satisfactory equation for the eigenfrequencies of the pairing vibrations, we use the expression for $B(\omega_{PV})$ (31) but with the following prescription: the semiclassical spectrum has a gap near the Fermi energy defined as

$$|\epsilon-\epsilon_F|_{\min} = d \tag{32}$$

with $d/\Delta \ll 1$. In this way we take into account the effect of the distribution of the discrete single-particle levels near the Fermi energy on the parameter $B(\omega_{PV})$. Then we get

$$B(\omega_{PV}) = \frac{\hbar^2}{4}\left[\int_0^{\mu-d} d\epsilon\,\tilde{g}(\epsilon)\frac{1}{E(\epsilon)(\epsilon-\mu)^2} + \int_{\mu+d}^\infty d\epsilon\,\tilde{g}(\epsilon)\frac{1}{E(\epsilon)(\epsilon-\mu)^2}\right] \approx \hbar^2 \tilde{g}(\mu)\frac{1}{2\Delta\,d}(1-\frac{d}{\Delta}) \tag{33}$$

and the expression for the relative spectroscopic factor (22) in the form

$$\frac{S_{PV}(p,t)}{S_0(p,t)} \approx \frac{d}{\tilde{g}(\mu)\,\Delta^2}(1-\frac{d}{\Delta})^{-1}. \tag{34}$$

The value of the relative spectroscopic factor for the excitation of monopole pairing vibrations in the two-neutron transfer reactions strongly depends on the distribution of the single-particle energy levels near the Fermi energy. The spectroscopic factor is proportional to the size of the gap near the Fermi energy. Numerical estimates of the relative spectroscopic factor based on (34) give value of order 0.06 for superfluid heavy nuclei. We used the usually employed pairing gap value in heavy nuclei $\Delta = 12/A^{1/2}$ MeV [20] and the pair-neutron level density for a spherical square- well potential of radius $R$, see Eq. (16). The gap parameter near the Fermi energy was chosen equal to d = 0.1 MeV. Our estimate is in agreement with the experimental relative cross sections $\sigma/\sigma_0$ for the excitation of $0^+$ states by the (p,t) reaction in the energy region of double pairing gap in even superfluid nuclei of the rare-earth and actinide regions. In particular, in $^{158}$Gd nucleus the experiment gives $\sigma/\sigma_0$=0.03 for the $0^+$ state with the energy E=1.957 MeV [1]; in $^{232}$U nucleus - $\sigma/\sigma_0$=0.02 for the $0^+$ state with the energy E=1.569 MeV [2]; in $^{228}$Th nucleus - $\sigma/\sigma_0$=0.06 for the $0^+$ state with the energy E=1.627 MeV [3].

## 4. Conclusions

The present semiclassical model of monopole pairing vibrations describes collective pairing excitations as explicitly related to dynamic fluctuations of the pairing gap in superfluid nuclei. The anomalous density response function is used to study of pairing collective effects. This response function associated with the self-consistent pairing-field fluctuations is appropriate for considering the excitation of monopole pairing vibrations in a two-neutron transfer reaction.

The semiclassical spectroscopic factor found in this paper contains coherence for the excitation of pairing vibrations in the two-neutron transfer reaction. This coherence is strongly dependent on the distribution of neutron levels near the Fermi surface. This distribution is determined by quantum effects, but found semiclassical spectroscopic factor does describe the average behavior of the intensity of the excitation of pairing vibrations in a two-neutron transfer reaction. The semiclassical spectroscopic factor allows us to estimate the observed cross sections for the two-neutron transfer to excited $0^+$ states in the energy region of the double pairing gap.

Studying of collective pairing effects in superfluid nuclei within the present semiclassical model is only the first step. Further extension of this approach, in particular, including fluctuations of the normal mean field is of great interest.

*This work was partially suppoted by the budget program "Support for the develement of priority areas of scientific researches" (code 6541230).*

REFERENCES

1. A.I. Levon, D. Bucurescu, C. Costache, T. Faestermann, R. Hertenberger, A. Ionescu, R. Lica, A.G. Magner, C. Mihal, R. Mihal, C.R. Nita, S. Pascu, K.P. Shevchenko, A.A. Shevchuk, A.

Turturica, H.F. Wirth. Phys. Rev. C 100 (2019) 034307.

2. A.I. Levon, P. Alexa, G. Graw, R. Hertenberger, S. Pascu, P.G. Thirolf, H.F. Wirth. Phys.Rev. C 92 (2015) 064319.
3. A.I. Levon, G. Graw, R. Hertenberger, S. Pascu, P.G. Thirolf, H.F. Wirth, and P. Alexa. Phys. Rev.C 88 (2013) 014310.
4. R. A. Broglia, O. Hansen, and C. Riedel, Adv. Nucl. Phys.6 (1973) 287.
5. Von Oertzen W., Vitturi A. Rep. Progr. Phys. 2001. V. 64. P. 1247.
6. D. R. Bes and R. Broglia, Nucl. Phys.80 (1966) 289.
7. Avez B., Simenel C., Chomas Ph. Phys. Rev. C. 78 (2008) 044318.
8. Shimoyama H., Matsuo M. Phys. Rev. C. 84 (2011) 044317.
9. Grasso M., Lacroix D., Vitturi A. Phys. Rev. C. 85 (2012) 034317.
10. D. Gambacurta, and D. Lacroix. Phys. Rev. C 86 (2012) 064320.
11. Assié M., Dasso C. H., Liotta, R. J. Macchiavelli A. O., and Vitturi A. eprint arXiv:1905.01339 (2019).
12. D.M. Brink, R.A. Broglia Nuclear Superfluidity: Pairing in Finite Systems. Cambridge University Press, UK, 2005.
13. Broglia, R.A., Zelevinsky, V. (eds.). Fifty Years of Nuclear BCS: Pairing in Finite Systems. World Scientific Publishing Co. Pte. Ltd., Singapore, 2013.
14. Abrosimov V.I., Brink D.M., Dellafiore A., Matera F. Nucl. Phys. A 864 (2011) 38.
15. Abrosimov V. I., Brink D. M., Matera F. Bull. Russ. Acad. Sci. 78 (2014) 630.
16. Abrosimov V.I., Brink D.M., Dellafiore A., Matera F. Nucl. Phys. A 800 (2008) 1.
17. Ring P., Schuck P. The nuclear many-body problem. Springer-Verlag, New York, 1980.
18. Strutinsky V.M., and Abrosimov V.I. Z. Phys. A 289 (1978) 83.
19. Cusson R.Y., and Hara K. Z. Phys. A 209 (1968) 428.
20. H. Olofsson, S. Åberg, and P. Leboeuf. Phys. Rev. Lett. 100 (2008) 037005.